\documentclass[11pt,a4paper,epsf,epsfig,psfrag]{article}
\usepackage{jheppub}
\usepackage{amsmath,graphicx,amsfonts, mathrsfs,amssymb}
\usepackage{amsmath,epsfig}
\usepackage{amssymb,amsfonts}
\usepackage{color}
\usepackage{latexsym}
\usepackage{epsfig}
\usepackage{pdfsync}
\newbox\pippobox

\def\be{\begin{equation}}
\def\ee{\end{equation}}
\def\bea{\begin{eqnarray}}
\def\eea{\end{eqnarray}}

\newcommand{\beq}{\begin{equation}}
\newcommand{\eeq}{\end{equation}}
\newcommand{\beqa}{\begin{eqnarray}}
\newcommand{\eeqa}{\end{eqnarray}}
\newcommand{\beqar}{\begin{eqnarray*}}
\newcommand{\eeqar}{\end{eqnarray*}}

\renewcommand{\eqref}[1]{(\ref{#1})}

\catcode`\@=12

\title{Inverse Magnetic Catalysis in the Soft-Wall Model of AdS/QCD }
\author[a]{Danning Li,}
\author[b,c]{Mei Huang,}
\author[d]{Yi Yang,}
\author[e]{Pei-Hung,Yuan}

\affiliation[a]{Department of Physics, Jinan University, Guangzhou 510632, P.R. China}
\affiliation[b]{Institute of High Energy Physics, Chinese Academy of Sciences, Beijing 100049, P.R. China}
\affiliation[c]{Theoretical Physics Center for Science Facilities, Chinese Academy of Sciences, Beijing 100049, P.R. China}
\affiliation[d]{Department of Electrophysics, National Chiao Tung University, Hsinchu, ROC}
\affiliation[e]{Institute of Physics, National Chiao Tung University, Hsinchu, ROC}

\emailAdd{lidanning@jnu.edu.cn}\emailAdd{huangm@ihep.ac.cn}\emailAdd{yiyang@mail.nctu.edu.tw}\emailAdd{phy.pro.phy@gmail.com}

\abstract{Magnetic effects on chiral phase transition have been investigated in a modified soft-wall AdS/QCD model, in which the dilaton field is taken to be negative at the ultraviolet region and positive at the infrared region as in Phys.Rev.D93(2016),101901 and JHEP1604(2016)036. The magnetic field is introduced into the background geometry by solving the Einstein-Maxwell system. After embedding the magnetized background geometry into the modified soft-wall model, the magnetic field dependent behavior of chiral condensate is worked out numerically. It is found that, in the chiral limit, the chiral phase transition remains as a second order at finite magnetic field $B$, while the symmetry restoration temperature and chiral condensate decrease with the increasing of magnetic field  in small $B$ region. When including finite quark mass effect, the phase transition turns to be a crossover one, and the transition temperature still decreases with increasing magnetic field $B$ when $B$ is not very large. In this sense, inverse magnetic catalysis effect is observed in this modified soft-wall AdS/QCD model.}

\begin{document}
\maketitle

\section{Introduction}
\label{sec-int}

Strong magnetic fields play essential roles in various physical systems, such as in the strong and weak phase transition of the universe \cite{Vachaspati:1991nm,Enqvist:1993np} and in noncentral heavy ion collisions at the Relativistic Heavy Ion Collider (RHIC) and the Large Hadron Collider (LHC)\cite{Skokov:2009qp,Voronyuk:2011jd,Bzdak:2011yy,Deng:2012pc}. Besides the phenomenological importance, in theoretical aspects of strong interaction, strong magnetic fields also provide good probe of the dynamics of quantum chromodynamics(QCD), of which the vacuum structure are of numerous interests.

Spontaneous chiral symmetry breaking, characterized by the non-zero quark condensate $\sigma\equiv \langle\bar{\psi}\psi\rangle$, is one of the most representative features of QCD vacuum. At sufficient high temperature, it is believed that chiral symmetry would be restored when chiral condensate is destroyed by the temperature effect. To speculate the effect of magnetic fields on the behavior of chiral condensate in QCD at both zero and finite temperatures is of great interest for decades. Since the 1990's, enhancement of quark condensate and increasing of transition temperature under magnetic fields, known as the magnetic catalysis, has been recognized \cite{Klevansky:1989vi,Klimenko:1990rh,Gusynin:1995nb,Shovkovy:2012zn}. The latter studies from effective models and approximations to QCD \cite{Shushpanov:1997sf, Agasian:1999sx, Alexandre:2000yf,
Agasian:2001hv, Cohen:2007bt, Gatto:2010qs, Gatto:2010pt, Mizher:2010zb, Kashiwa:2011js,
Avancini:2012ee, Andersen:2012dz, Scherer:2012nn}, as well as lattice simulations  \cite{Buividovich:2008wf,
Braguta:2010ej, D'Elia:2010nq, D'Elia:2011zu,Ilgenfritz:2012fw} gave results in coincidence with these studies. Nevertheless, an inverse result showing that the transition temperature decreases with increasing magnetic field, known as the inverse magnetic catalysis(IMC), has just been discovered in recent lattice simulation \cite{Bali:2011qj,Bali:2012zg}. This surprising result is confirmed by latter lattice simulations \cite{Bornyakov:2013eya}. Several efforts from different approaches \cite{Fukushima:2012kc, Kojo:2012js, Bruckmann:2013oba, Chao:2013qpa, Fraga:2013ova, Ferreira:2014kpa, Farias:2014eca, Yu:2014sla, Andersen:2014oaa, Ferrer:2014qka, Feng:2014bpa} have been made to explain the microscopic mechanism of this phenomena.

In QCD, it is well known that the dominant physics related to chiral phase transition is non-perturbative, when the normal perturbative methods become invalid. Besides lattice simulation, the discovery of the anti-de Sitter/conformal field theory (AdS/CFT) correspondence and the conjecture of the gravity/gauge duality \cite{Maldacena:1997re,Gubser:1998bc,Witten:1998qj} offers a new powerful tool to solve the strong coupling problem in the relevant physical region. In dealing with chiral phase transition with magnetic fields, from top-down holographic approach, a kind of IMC effect has been realized in \cite{Li:2016gtz}. Instead of calculating the order parameter of chiral phase transition, the authors considered the phase transition by comparing the free energy of different geometric configuration. Similarly, \cite{Mamo:2015dea} tried to investigate the phase transition temperature by comparing the free energy of different geometric backgrounds in bottom-up approach. They also found that the transition temperature decreases with increasing magnetic field. In some sense, it is a kind of IMC for deconfinement phase transition, since the geometric transition in bottom-up approach is usually considered as confinement/deconfinement phase transition. In other ways, the authors of \cite{Evans:2016jzo} proposed a single scalar model coupled with a $U(1)$ gauge field. After inserting the anomalous dimension into the scalar field, they observed the inverse magnetic catalysis behavior from chiral condensate, the order parameter of chiral phase transition.

Recently, from a more realistic bottom-up holographic framework, \cite{Dudal:2015wfn} tried to investigate the IMC effect within the soft-wall framework  \cite{Karch:2006pv},  which has been successfully applied and tested in describing hadron physics and relevant phenomena \cite{Gherghetta-Kapusta-Kelley,Gherghetta-Kapusta-Kelley-2,YLWu,YLWu-1,Fang:2016uer,Li:2012ay,Li:2013oda,Colangelo:2008us,He:2013qq,Chen:2015zhh,Capossoli:2016ydo,Capossoli:2016kcr}. However, by simply extending the original soft-wall model and considering the magnetized geometric background from Einstein-Maxwell system\cite{D'Hoker:2009bc,D'Hoker:2009mm}, the magnetic field dependent chiral condensate showed that there is no IMC effect in the original soft-wall model by this scenario. Nevertheless, as noted in \cite{Chelabi:2015cwn,Chelabi:2015gpc}, the background settings of original soft-wall model cannot describe spontaneous chiral symmetry breaking in the vacuum and its restoration at sufficient high temperature well. Thus, a modified version of soft-wall model was proposed in \cite{Chelabi:2015cwn,Chelabi:2015gpc}. After introducing a modified dilaton field and extending the soft-wall model to finite temperature, chiral phase transition are realized perfectly. Therefore, it is interesting to study the magnetic field effects on chiral phase transition in this modified soft-wall model.

In this work, we will extend the study in \cite{Chelabi:2015cwn,Chelabi:2015gpc}, and try to investigate how magnetic fields affect chiral condensate and chiral phase transition. The paper is organized as follows: in Sec.\ref{gravity}, we will give a brief introduction on how to introduce the magnetic field from the Einstein-Maxwell system. In Sec.\ref{magneticeff}, we extract the results of magnetic field dependent chiral condensate from the modified soft-wall model, both in the chiral limit and at finite quark mass. Finally, a short conclusion and discussion will be given in Sec.\ref{sum}.

\section{Gravity background}
\label{gravity}

In \cite{Chelabi:2015cwn,Chelabi:2015gpc},  the spontaneous chiral symmetry breaking in the vacuum and its restoration at finite temperature are realized perfectly in the soft-wall AdS/QCD model. There, the soft-wall model was put into the AdS-Schwarzchild(AdS-SW) black hole background as a probe. In these studies, only temperature effect and mass effect on chiral condensate are considered. To introduce the magnetic field, as was done in \cite{Mamo:2015dea,Dudal:2015wfn}, a simple way is to consider the gravity background with back-reaction of magnetic field through the Einstein-Maxwell(EM) system, with the action\footnote{It is more rigorous to introduce magnetic field by considering the relation between electric charge, baryon number and $I_3$, the last two of which are related to the global symmetry of QCD. Here, we just follow \cite{Mamo:2015dea,Dudal:2015wfn} and take the simplest version to introduce magnetic field. }
\begin{equation}
S_{B}= \frac{1}{16\pi G_5}\int d^5x\sqrt{-g}\big(R-F^{MN}F_{MN}+\frac{12}{L^2}),
\end{equation}
where $R$ is the scalar curvature, $G_5$ is the 5D Newton constant, $g$ is the determinant of metric $g_{\mu\nu}$, $F_{MN}$ is a $U(1)$ gauge field, and $L$ is the $AdS$ radius.

Within this system, the authors of \cite{Mamo:2015dea} compared the free energy of thermal AdS and black hole solution. In this way, they extracted the transition temperature and found that it would decrease with increasing magnetic field, showing a kind of inverse magnetic catalysis. As noted in the introduction, in some sense this kind of transition is related to confinement/deconfinement phase transition, since from the previous studies\cite{Gursoy:2008za,Colangelo:2010pe,Li:2011hp,Cai:2012xh,Li:2014hja,Li:2014dsa,Yang:2014bqa,Yang:2015aia,Fang:2015ytf} the relevant order parameters to distinguish the two phases are the loop operators, which is related to confinement/deconfinement phase transition more closely. To be more concrete and limited in chiral phase transition, Ref.\cite{Dudal:2015wfn} put the soft-wall model into the same metric background and extracted the magnetic dependent behavior of chiral condensate, which is the order parameter of chiral phase transition. They found that there is no inverse magnetic catalysis for chiral phase transition using this scenario. However, the simple version of soft-wall model cannot give well description of spontaneous chiral symmetry breaking even in the vacuum\cite{Chelabi:2015cwn,Chelabi:2015gpc}. Thus, it is necessary to consider the magnetic effects in an improved version of soft-wall model. For simplicity, we will first consider the simple gravity background given by the Einstein-Maxwell system. The back-reaction from matter part(soft-wall action) to the metric will be neglected in this simple version. Before we move to deal with chiral condensate, in this section, we will first give a short introduction about the background gravity solution for preparation.

\subsection{Equation of Motion in EM system}

The Einstein equation for the EM system could be easily found in the text books, and it takes the form
\begin{equation}
E_{M N}-\frac{6}{L^2}g_{M N}-2(g^{IJ} F_{MI}F_{NJ}-\frac{1}{4}F_{IJ}F^{IJ}g_{MN})=0,
\end{equation}
with $E_{M N}$ the Einstein tensor, defined in terms of the Ricci tensor $R_{MN}$ and Ricci scalar $R$ as $E_{MN}=R_{MN}-\frac{1}{2}R~g_{MN}$.

The field equation for $F_{MN}$ is of the form
\begin{equation}
\nabla_M F^{MN}=0,
\end{equation}
and one can check that the constant magnetic field configuration
\begin{equation}\label{config-F}
F=\frac{B}{L} dx_1\wedge dx_2
\end{equation}
satisfies this equation.

Since the magnetic field along $x_3$ axis breaks the rotation symmetry, the ansatz for the metric could be taken as
\begin{equation}
dS^2=\frac{L^2}{z^2}\left(-f(z)dt^2+\frac{1}{f(z)}dz^2+h(z)(dx_1^2+dx_2^2)+q(z) dx_3^2\right).
\end{equation}
For a black hole solution, $f(z=z_h)=0$ at horizon $z=z_h$ and  $q(z)$ together with $h(z)$ are regular function of $z$ in the region $0<z<z_h$.

Under this metric ansatz and the magnetic field configuration Eq.(\ref{config-F}), the Einstein equations could be simplified as
\begin{equation}\label{eq-f}
f''+ \left(\frac{h'}{3 h}+\frac{q'}{6 q}-\frac{3}{z}\right)f'+\left(-\frac{2 h' q'}{3 h q}-\frac{h'^2}{3 h^2}+\frac{2 h'}{z h}+\frac{q'}{z q}\right)f-\frac{8 B ^2 z^2}{3 h^2}=0,
\end{equation}

\begin{equation}\label{eq-q}
q''+\left(\frac {2 f'} {3 f}+\frac {h'}{3 h}-\frac{2}{z}\right)q'+\left(-\frac {2 f' h'}{3 f h}-\frac {h'^2}{3 h^2}+\frac {2 h'}{z h}-\frac{8 B^2 z^2}{3 f h^2}\right)q-\frac{q'^2}{2 q}=0,
\end{equation}

and
\begin{equation}\label{eq-h}
h''+\left(\frac {f'}{3 f}-\frac {q'}{6 q}-\frac{1}{z}\right) h' + \left(-\frac{f' q'}{3 f q}+\frac{q'}{z q}\right)h-\frac {h'^2}{3 h}+\frac{4 B^2 z^2}{3 f h}=0,
\end{equation}

together with a constrain equation
\begin{equation}\label{eq-c}
\frac{f' h'}{2 f h}+\frac{f' q'}{4 f q}+\frac{h' q'}{2 h q}+\frac {h'^2}{4 h^2}-\frac {3 f'}{2 z f}-\frac{3 h'}{z h}-\frac{3 q'}{2 z q}-\frac {6} {z^2 f}+\frac{6}{z^2}+\frac{B^2 z^2}{f h^2}=0.
\end{equation}

The next subsection will be devoted to solve these equations and extract the metric background with magnetic field.

\subsection{Perturbative solution}

Requiring asymptotic AdS boundary condition, it is hard to get the exact analytical solution of Eqs.(\ref{eq-f}-\ref{eq-c}). The full solution requires numerical method. Nevertheless, as given in \cite{D'Hoker:2009bc}, the leading expansion for these equations could be easily derived, taking the form
\begin{eqnarray}
f(z)=&&1+\frac{2}{3} B^2\ln(z/L)z^4+ f_4 z^4,\label{f4L}\\
q(z)=&&1+\frac{2}{3} B^2 \ln(z/L)z^4,\label{q4L}\\
h(z)=&&1-\frac{1}{3} B^2 z^4 \ln(z/L),\label{h4L}
\end{eqnarray}
with $f_4$ the integral constants related to the black hole horizon $z_h$, where$f(z)=0$. Here, as pointed out in \cite{D'Hoker:2009mm}, $B$ is related with the physical magnetic field  $\mathcal{B}$ at the boundary by the equation $\mathcal{B}=\sqrt{3} B$. Latter, for simplicity, we will consider $B$ instead of $\mathcal{B}$. We will see that the expansion could be a good approximation in the small $B$ and high temperature $T$ region. One can check that when $B$ goes to zero, it would reduce to the AdS-SW black hole solution. Thus, if the temperature and magnetic filed of interest are in this region, we can use this expansion as an approximation. Indeed, in \cite{Mamo:2015dea,Dudal:2015wfn}, the authors took this leading expansion as the metric background and studied the magnetic effects on the phase transition temperature. Since they only considered small magnetic field and in the final results, the relevant transition temperatures are not very low(in the relevant magnetic field region), the qualitative results are reliable. In this sense, because we are interested only in the region of near transition point and focus only in the qualitative results, we can follow these studies and only consider the perturbative expansion as first approximation.
%

Before we start to study the behavior of chiral condensate under magnetic fields, we would like to make more efforts on estimating the errors by using the perturbative solution Eqs.(\ref{f4L}-\ref{h4L}). Firstly, we try to get perturbative solutions of Eqs.(\ref{eq-f}-\ref{eq-c}) with more powers of $z$. Here, as an example, we list the solutions up to $z^{12}$. One can check the following near boundary expansion could satisfy Eqs.(\ref{eq-f}-\ref{eq-c}) to $O(z^{13})$

\begin{eqnarray}
f^{(12)}(z)=&&1+\frac{2}{3} B^2\ln(\mu z)z^4 + f_4 z^4\nonumber\\
&&+B^4 \left(-\frac{2}{63}\ln ^2(\mu  z)+\frac{25}{441} \ln (\mu  z)-\frac{181}{8232}\right)z^8\nonumber\\
&&+B^4 f_4 \left(\frac{5}{231} \ln ^2(\mu  z)-\frac{1003}{106722}\ln (\mu  z)+\frac{209015}{131481504}\right) z^{12}\nonumber\\
&&+B^6 \left(\frac{10}{693} \ln ^3(\mu  z)-\frac{2773}{640332} \ln ^2(\mu  z)+\frac{626341 }{197222256}\ln (\mu  z)-\frac{12782437}{17355558528}\right)z^{12}\nonumber\\
&&+O(z^{16})
\end{eqnarray}

\begin{eqnarray}
q^{(12)}(z)=&&1+\frac{2}{3} B^2 \ln(\mu  z)z^4 \nonumber\\
&&-B^2 f_4 \left(\frac{1}{3}  \ln (\mu  z)+\frac{1}{24} \right)z^8+B^4 \left(-\frac{2}{63}\ln ^2(\mu  z)+\frac{25}{441} \ln (\mu  z)-\frac{181}{8232}\right)z^8\nonumber\\
&&+B^2 f_4^2 \left(\frac{2}{9}\ln (\mu  z)+\frac{1}{27}\right)z^{12}+B^4 f_4 \left(\frac{10}{99} \ln ^2(\mu  z)-\frac{148 }{3267}\ln (\mu  z)+\frac{20051}{1724976}\right)z^{12}\nonumber\\
&&+B^6 \left(\frac{10}{693} \ln ^3(\mu  z)-\frac{2773 }{640332}\ln ^2(\mu  z)+\frac{626341 }{197222256}\ln (\mu  z)-\frac{12782437}{17355558528}\right) z^{12}\nonumber\\
&&+O(z^{16})
\end{eqnarray}

\begin{eqnarray}
h^{(12)}(z)=&&1-\frac{1}{3} B^2 z^4 \ln(\mu  z) \nonumber\\
&& B^2 f_4 \left(\frac{1}{6} \ln (\mu  z)+\frac{1}{48}\right)z^8+B^4 \left(\frac{17}{126} \ln ^2(\mu  z)-\frac{47}{1764}\ln (\mu  z)+\frac{305}{32928}\right)z^8\nonumber\\
&&- B^2 f_4^2 \left(\frac{1}{9} \ln (\mu  z)+\frac{1}{54}\right)z^{12}+B^4 f_4 \left(-\frac{35}{198}  \ln ^2(\mu  z)+\frac{415 }{52272}\ln (\mu  z)-\frac{17131}{3449952}\right)z^{12}\nonumber\\
&&+ B^6 \left(-\frac{101 }{1386}\ln ^3(\mu  z)+\frac{28939 }{1280664}\ln ^2(\mu  z)-\frac{590707 }{65740752}\ln (\mu  z)+\frac{38983919}{104133351168}\right)z^{12}\nonumber\\
&&+O(z^{16})
\end{eqnarray}

In the above expressions, we have two integral constants $f_4$ and $\mu$. If one takes $\mu=\frac{1}{L}$, then the leading expansion would be exactly the one in Eqs.(\ref{f4L}-\ref{h4L}). However, for a black hole solutions, these two integral constants are not independent, due to the near horizon boundary condition. Usually, the near horizon boundary condition are required as the regular condition of the equation of motion. The definition of the horizon $z_h$ is $f(z_h)=0$, so in Eqs.(\ref{eq-q},\ref{eq-h}) apparent singularity appears where $f(z)$ arises as the denominators. If we collect such kind of terms out, we find that in Eq.(\ref{eq-q}) and Eq.(\ref{eq-h}) they takes the form $\frac{Q(z)}{f(z)}$ and $-2\frac{q(z)}{h(z)}\frac{Q(z)}{f(z)}$, with
\begin{equation}
Q(z)\equiv-\frac{8B^2z^2q}{3h^2}-\frac{2 q f' h'}{3 h}+\frac{2}{3} f' q'.
\end{equation}
Since we require the solutions to be regular at horizon, there would be a natural regular condition
\begin{equation}\label{IR-boundary}
Q(z_h)=0
\end{equation}
to cancel the $f(z_h)=0$ singularity at horizon. This regular condition would require $\mu$ to depend on $f_4$. In this way, there are no free parameters in the full solutions. In general, when $B$ is fixed, $\mu,f_4$ would be a function of temperature $T$. Thus, the full solutions would be fixed when $B, T$ are given.

\begin{figure}[h]
\begin{center}
\epsfxsize=4.5 cm \epsfysize=4.5 cm \epsfbox{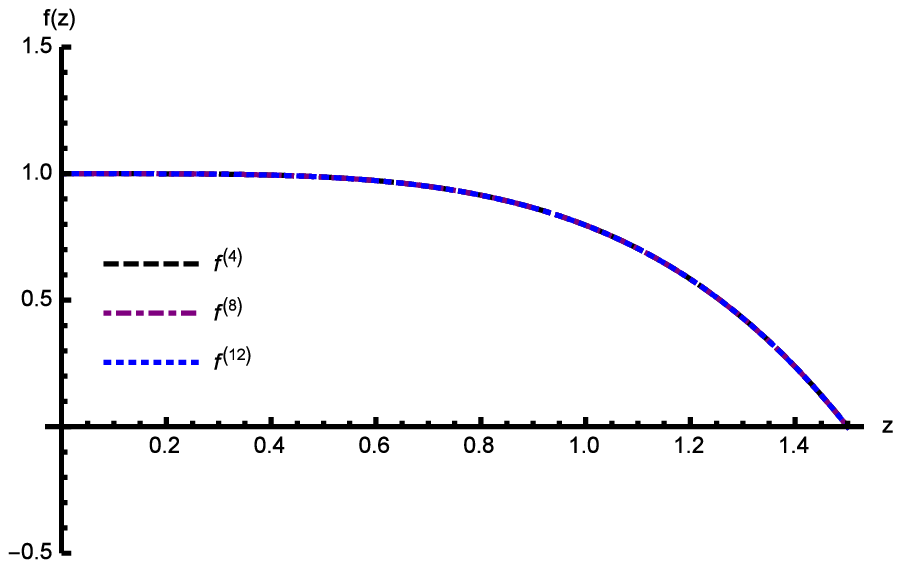}
\hspace*{0.1cm}
\epsfxsize=4.5 cm \epsfysize=4.5 cm\epsfbox{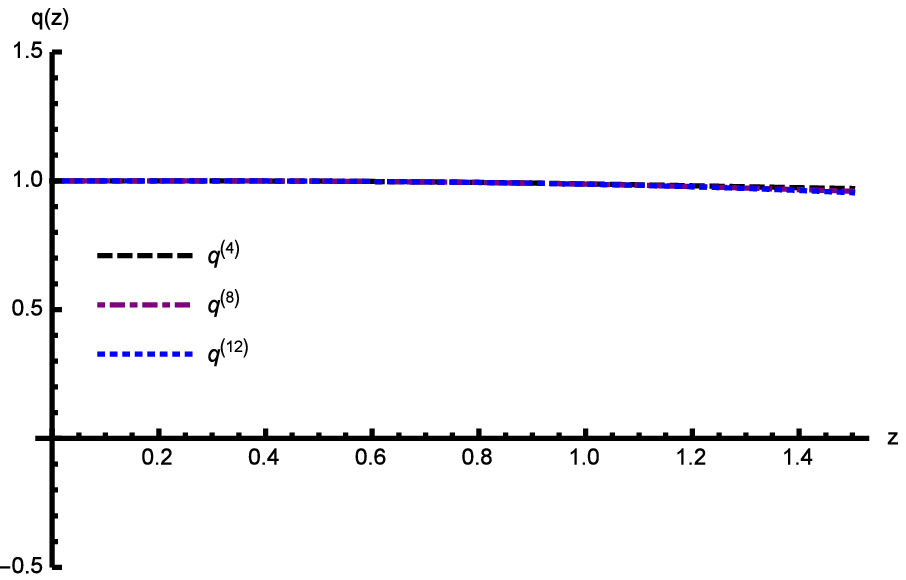}
\hspace*{0.1cm}
\epsfxsize=4.5 cm \epsfysize=4.5 cm\epsfbox{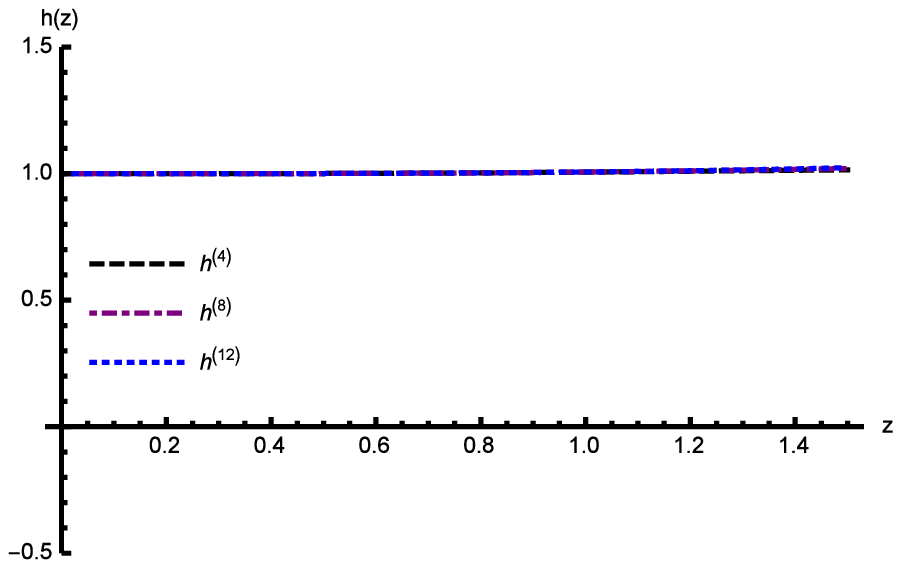}
\vskip -0.05cm
\hskip 0.15 cm
\textbf{ (a)  } \hskip 4.5 cm \textbf{(b) } \hskip 4.0 cm \textbf{(c) }\\
\end{center}
\caption{Different orders of functions $f(z),q(z),h(z)$ as functions of $z$ when $B=0.15{\rm GeV}^2,\mu=0.45 {\rm GeV},z_h\simeq1.5{\rm GeV}^{-1}$, which gives $T\simeq 208{\rm MeV}$.}
 \label{diff-fqh-s}
\end{figure}

\begin{figure}[h]
\begin{center}
\epsfxsize=4.5 cm \epsfysize=4.5 cm \epsfbox{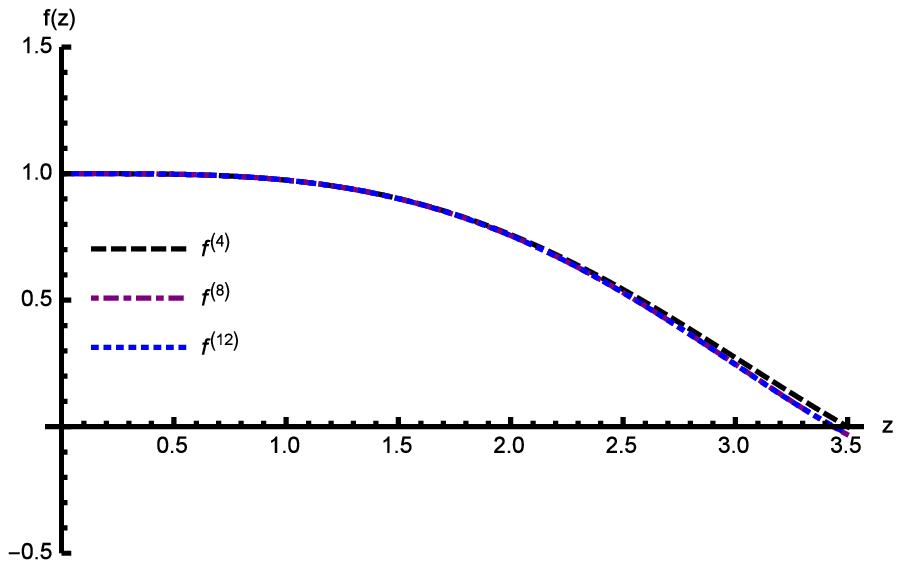}
\hspace*{0.1cm}
\epsfxsize=4.5 cm \epsfysize=4.5 cm\epsfbox{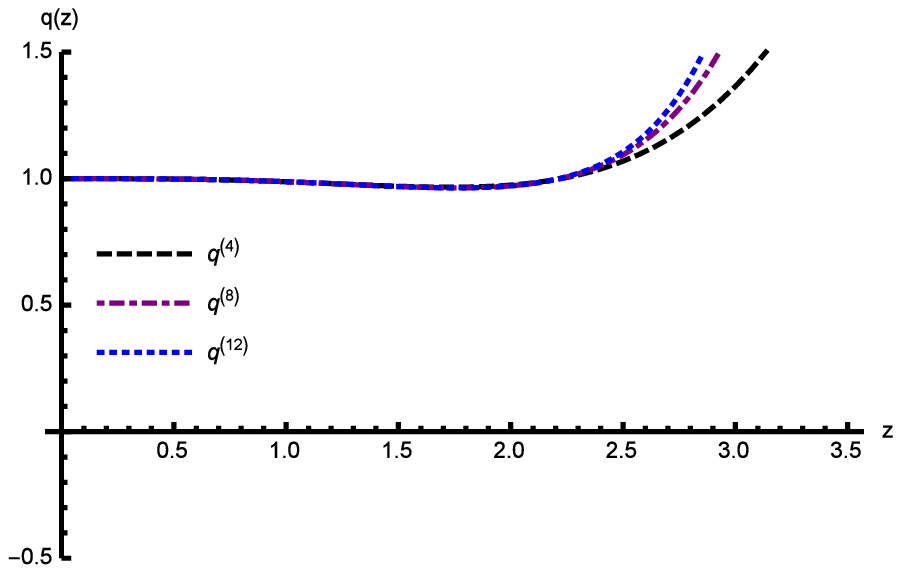}
\hspace*{0.1cm}
\epsfxsize=4.5 cm \epsfysize=4.5 cm\epsfbox{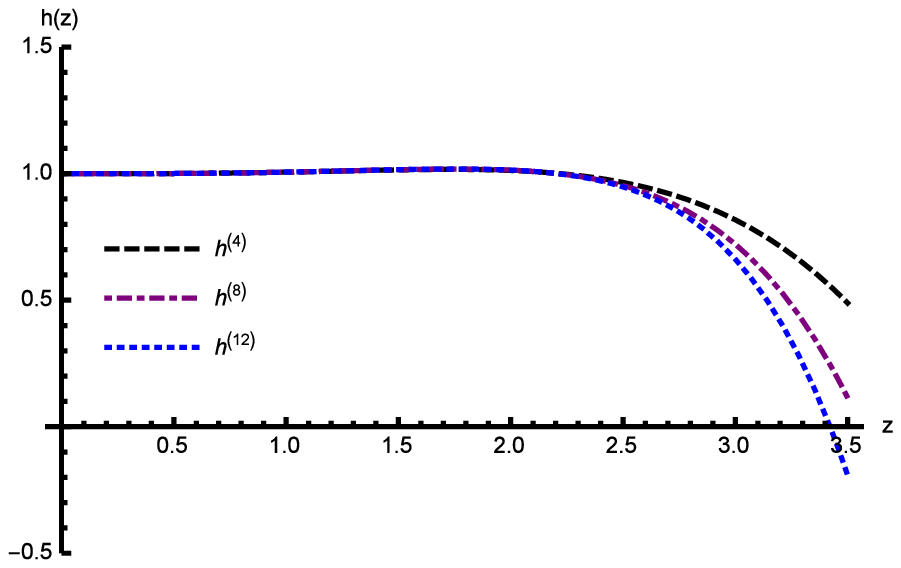}
\vskip -0.05cm
\hskip 0.15 cm
\textbf{ (a)  } \hskip 4.5 cm \textbf{(b) } \hskip 4.0 cm \textbf{(c) }\\
\end{center}
\caption{Different orders of functions $f(z),q(z),h(z)$ as functions of $z$ when $B=0.15{\rm GeV}^2,\mu=0.45 {\rm GeV},z_h\simeq3.5{\rm GeV}^{-1}$, which gives $T\simeq 40{\rm MeV}$.}
 \label{diff-fqh-l}
\end{figure}

From the structure of the UV expansion, the leading, next-leading, next-next-leading powers are of $z^4$, $z^8$, $z^{12}$ respectively. In between those terms, there are no powers like $z^5,z^6,z^7, z^9,z^{10},z^{11}$. We would expect that the leading expansion could provide good approximation in small $z$ area. Thus, if $z_h$ is very small, we could expect that the perturbative solution could be used in the whole region from boundary to horizon. As pointed out in \cite{Mamo:2015dea}, the perturbative solution could work well when $B<<T^2$. To be more explicitly, we plot the functions $f^{(4)},q^{(4)},h^{(4)},f^{(8)},q^{(8)},h^{(8)},f^{(12)},q^{(12)},h^{(12)}$ to show the errors for small $T$ and large $T$ in Fig.\ref{diff-fqh-s} and Fig.\ref{diff-fqh-l}. In Fig.\ref{diff-fqh-s}, we take $B=0.15{\rm GeV}^2,\mu=0.45 {\rm GeV},z_h\simeq1.5{\rm GeV}^{-1}$, which gives $T\simeq 208{\rm MeV}$. We could see that in Panel(a),(b),(c), the three lines for perturbative expansions to different orders are almost the same, showing good convergence of the perturbative expansion in this value of magnetic field and temperature. Simultaneously, in Fig.\ref{diff-fqh-l}, we plot the corresponding functions when $B=0.15{\rm GeV}^2,\mu=0.45 {\rm GeV},z_h\simeq3.5{\rm GeV}^{-1}$ and $T\simeq 40{\rm MeV}$. There we could see that the convergence of the perturbative expansion becomes very bad. $h^{(12)}(z)$ is even negative within $0<z<z_h$, showing the invalidity of the expansion solution in this physical region, where a full numerical solution is necessary.

Thus, if considering a temperature region not very low, we can take the leading expansion as a first test to catch the qualitative behavior, instead of trying to solve the full metric background. From the above discussion, we can expect that the small $B$ and large temperature results would be reliable, since the expansion solution could be good approximation to the full solutions in this corner. As a first approximation, we will take the leading expansion
\begin{eqnarray}
f(z)=&&1-\frac{z^4}{z_h^4}\left(1+\frac{2}{3}B^2\log(\frac{z}{z_h})\right),\label{f4L1}\\
q(z)=&&1+\frac{2}{3} B^2 \ln(\mu z)z^4,\label{q4L1}\\
h(z)=&&1-\frac{1}{3} B^2 z^4 \ln(\mu z),\label{h4L1}
\end{eqnarray}
To be more careful, as in \cite{Mamo:2015dea,Dudal:2015wfn}, we will vary $\mu$ from $0.25\rm{GeV}$ to $1\rm{GeV}$ to check the convergence of the expansion and estimate the possible errors in the later calculation\footnote{But we emphasize that $\mu$ is not a free parameter. Instead, in principle, it could be solved by considering the near horizon boundary condition. However, when the perturbative expansion are valid, the difference of solutions with different $\mu$ is very small.}.

\section{Magnetic field effects on chiral phase transition}
\label{magneticeff}

In the previous section, we have introduced the magnetic field into the metric background by solving the EM system. In this section, we will take the perturbative expansion Eqs.(\ref{f4L1}-\ref{h4L1}) as the background geometric and consider its effect on chiral phase transition. We will extend the studies in \cite{Chelabi:2015cwn,Chelabi:2015gpc} to finite $B$ case and try to extract the magnetic effects on chiral phase transition in the framework holography. More concretely, we will embed the following soft-wall action in this perturbative metric background:
\begin{eqnarray}\label{kkssaction}
 S=&&-\int d^5x
 \sqrt{-g}e^{-\Phi}Tr(D_m X^+ D^m X+V_X+\frac{1}{4g_5^2}(F_L^2+F_R^2)),
\end{eqnarray}
with $A_{L/R}$ the left/right hand gauge field, $D_m$ the covariant derivative defined as $D_mX=\partial_mX-i A^L_mX+i XA^R_m$, $F_{mn}$ the field strength defined as $F_{mn}=\partial_m A_{n}-\partial_n A_{m}-i[A_m,A_n]$, $V_X$ the scalar potential, $g_5$ the gauge coupling, $g$ the determinant of the metric, and $\Phi$ the dilaton field. The leading term of $V_X$ is the mass term of the complex scalar field $X$ $M_5^2 X X^+$ and $M_5^2$ can be determined as $M_5^2=-3$(we take the AdS radius $L=1$ in this work) from the AdS/CFT prescription $M_5^2=(\Delta-p)(\Delta+p-4)$\cite{Witten:1998qj} by taking $\Delta=3, p=0$.

As \cite{Dudal:2015wfn}, we consider the simplest case and the magnetic field are taken into account only through the metric. We would not consider the possibility of higher order coupling between the magnetic filed and $X$ like in \cite{Evans:2016jzo}. As in \cite{Karch:2006pv}, we assume that the vacuum expectation value of $X$ field takes the diagonal form $X=\frac{\chi}{2}I_2$ in $SU(2)$ case. Under the metric ansatz Eq.(\ref{eq-c}),  the equation of motion for $\chi$ would be

\begin{equation}\label{equation-chi}
\chi^{''}+(\frac{-3}{z}-\Phi^{'}+\frac{q^{'}}{2q}+\frac{h^{'}}{h}+\frac{f^{'}}{f})\chi^{'}-\frac{1}{z^2f}\partial_\chi V(\chi)=0,
\end{equation}
where we have defined $V(\chi)=Tr{V_X}$.

When $B=0$, we have $q=h\equiv1$ and the above equation reduces to the form
\begin{equation}\label{equation-chi-o}
\chi^{''}+(\frac{-3}{z}-\Phi^{'}+\frac{f^{'}}{f})\chi^{'}-\frac{1}{z^2f}\partial_\chi V(\chi)=0.
\end{equation}
just the same as the one without magnetic field as shown in \cite{Chelabi:2015cwn,Chelabi:2015gpc}. If we look into Eq.(\ref{equation-chi}), magnetic effects enter this equation through three metric functions $f,q,h$. Since $\Phi,q,h$ only appears in the $\chi^{'}$ terms, we can recombine them and define an effective dilaton $\phi=\Phi-\log(\sqrt{q} h)$. Using $\phi$, Eq.(\ref{equation-chi}) becomes
\begin{equation}\label{equation-chi-o1}
\chi^{''}+(\frac{-3}{z}-\phi^{'}+\frac{f^{'}}{f})\chi^{'}-\frac{1}{z^2f}\partial_\chi V(\chi)=0.
\end{equation}
In the numerical analysis of \cite{Chelabi:2015cwn,Chelabi:2015gpc}, it is found that a large negative part of dilaton field would enhance the chiral condensate (See Fig.8 and Fig.9 in \cite{Chelabi:2015cwn,Chelabi:2015gpc}). Therefore, if the extra term $-\log(\sqrt{q} h)$ is negative, we would expect a magnetic catalysis effects, while if it is positive we will expect the IMC. Since we only consider in the small $B$ region for the perturbative solution, we could expand $-\log(\sqrt{q} h)$ in powers of $B$, and the leading term is $z^8\log^2(\mu z)B^4>0$. In this sense, we would expect an IMC effect in small $B$ and large $T$ region, while the qualitative behavior in other region requires a full solution. The dilaton field could be dual to the coupling constant, in some sense, the magnetic field dependent effective dilaton field works like a coupling constant running with magnetic field. Thus, if there is IMC in this scenario, in some sense it supports the studies in Refs.\cite{Ferreira:2014kpa,Ferrer:2014qka}. But here, we emphasize that the above analysis is a quite rough one. We have not considered the changes of $f(z)$, so in the next sections we will check it numerically.

\subsection{Inverse magnetic catalysis in the chiral limit}

It is easy to check that the UV structure of the solution of $\chi$ still has the form
\begin{eqnarray}
\chi=m_q\zeta z+...+\frac{\sigma}{\zeta}z^3+...,
\end{eqnarray}
with $m_q,\sigma$ two integral constants of Eq.(\ref{equation-chi}), standing for quark mass and chiral condensate respectively, and $\zeta=\sqrt{3}/(2\pi)$ a normalization constant\cite{Cherman:2008eh}. As explained in \cite{Karch:2006pv,Chelabi:2015cwn,Chelabi:2015gpc}, there is an additional boundary condition $f^{'}\chi^{'}-e^{2A_s}\partial_\chi V(\chi)=0$ at the horizon $z=z_h$ where $f(z_h)=0$. This IR boundary condition would make $\sigma$ be a function of quark mass, temperature and magnetic field. Using the numerical method described in \cite{Chelabi:2015cwn,Chelabi:2015gpc}, we can extract $\sigma$ for given $m_q, T, B$ numerically.  As an extension of the previous work, we still take the dilaton of the form
\begin{eqnarray}\label{int-dilaton}
\Phi(z)=-\mu_1^2z^2+(\mu_1^2+\mu_0^2)z^2\tanh(\mu_2^2z^2),
\end{eqnarray}
which is of negative quadratic form $-\mu_1^2 z^2$ in the UV region and positive quadratic form $\mu_0^2 z^2$ in the IR region. Here, in principle, the dilaton field could depend on the magnetic filed. To determine the exact dependent behavior, we need to solve the Einstein-Dilaton-Maxwell coupled system, which is quite complicated. We will leave this hard task to the future and only consider the simple assumption in Eq.(\ref{int-dilaton}). The scalar potential $V(\chi)$ will also be kept as
\begin{eqnarray}
V(\chi)=-\frac{3}{2}\chi^2+v_4\chi^4.
\end{eqnarray}
$\mu_0$ is related to the Regge slope of light meson spectral and is fixed to be $\mu_0=0.43{\rm GeV}$. In \cite{Chelabi:2015cwn,Chelabi:2015gpc}, $\mu_1,\mu_2,v_4$ are fixed to be $v_4=8,\mu_1=0.830\rm{GeV},\mu_2=0.176\rm{GeV}$, with which the chiral phase transition temperature and the value of chiral condensate in vacuum are set to be around $150{\rm MeV}$ and $(327 {\rm MeV})^3$,  in coincidence with lattice results. In the latter calculation, we will stick to the same parameter values.

\begin{figure}[h]
\begin{center}
\epsfxsize=6.5 cm \epsfysize=6.5 cm \epsfbox{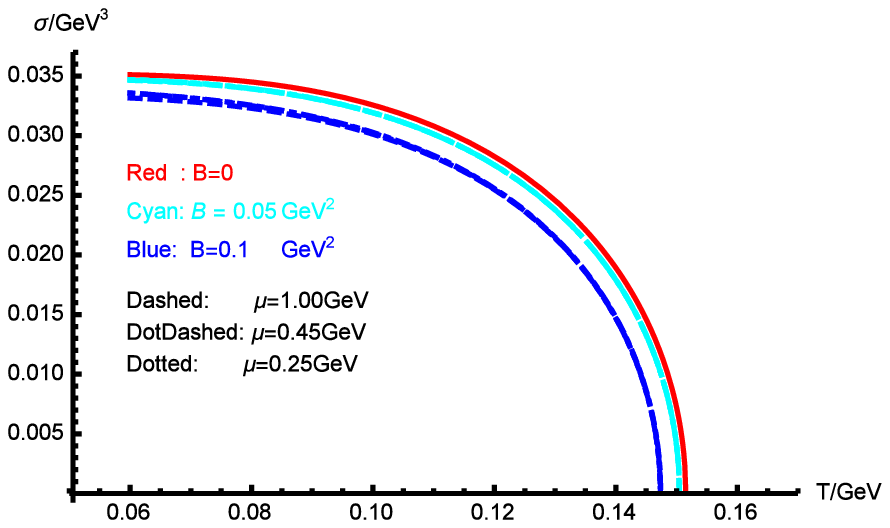}
\hspace*{0.1cm} \epsfxsize=6.5 cm \epsfysize=6.5 cm \epsfbox{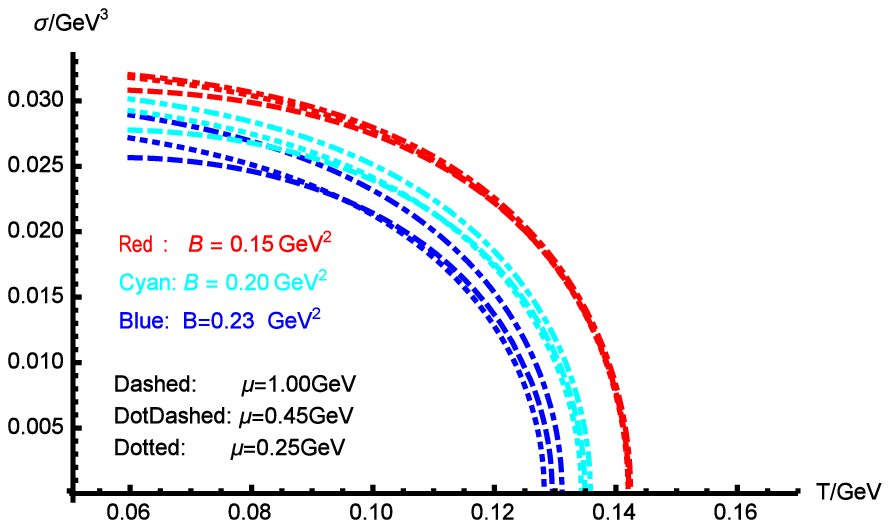}
\vskip -0.05cm \hskip 0.15 cm
\textbf{( a ) } \hskip 6.5 cm \textbf{( b )} \\
\end{center}
\caption{$\sigma$ as a function of the temperature $T$ for different values of magnetic field $B=0,0.05,0.1,0.15,0.2,0.23{\rm GeV}^2$, when $m_q=0, v_4=8,\mu_0=0.430\rm{GeV},\mu_1=0.830\rm{GeV},\mu_2=0.176\rm{GeV}$. Panel.(a) gives the results for $B=0,0.05,0.1{\rm GeV}^2$ in red, cyan, blue lines respectively. Panel.(b) gives the results for $B=0.15,0.2,0.23{\rm GeV}^2$ in red, cyan, blue lines respectively. Both in Panel.(a) and (b), the dashed, dotdashed, dotted lines represent $\mu=1.0,0.45,0.25 {\rm GeV}$ respectively. For $B=0$, there is no $\mu$ dependence, and it is plotted in solid red line.}
\label{sigma-B-T-m0}
\end{figure}



\begin{figure}[h]
\begin{center}
\epsfxsize=6.5 cm \epsfysize=6.5 cm \epsfbox{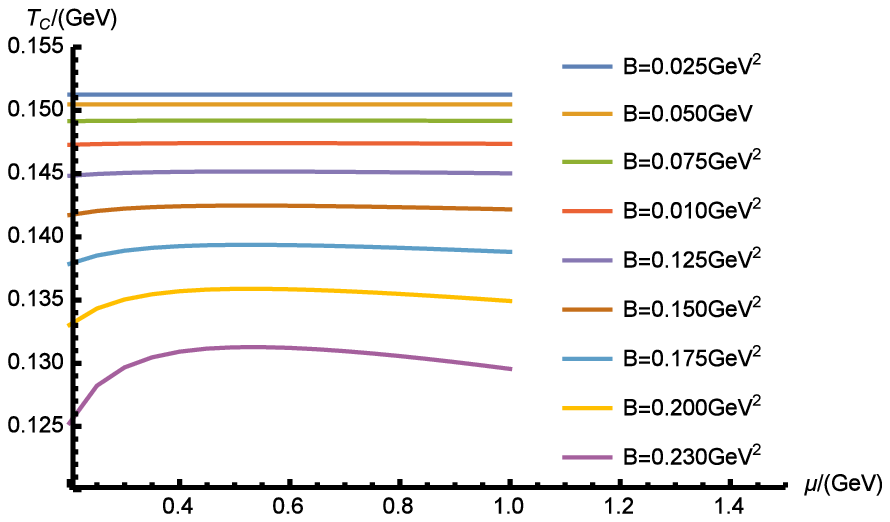}
\hspace*{0.1cm} \epsfxsize=6.5 cm \epsfysize=6.5 cm \epsfbox{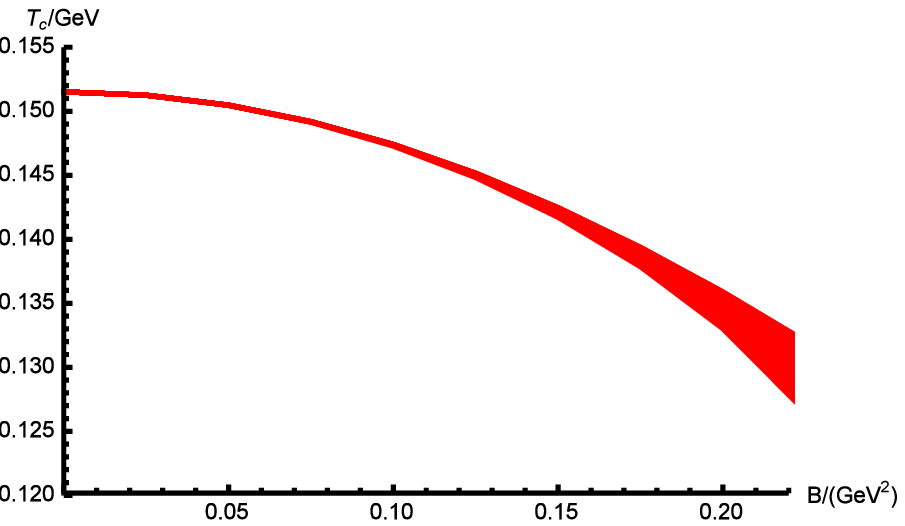}
\vskip -0.05cm \hskip 0.15 cm
\textbf{( a ) } \hskip 6.5 cm \textbf{( b )} \\
\end{center}
\caption{The critical temperature $T_c$ as a function of the magnetic field $B$ in the chiral limit $m_q=0$, when $v_4=8,\mu_0=0.430\rm{GeV},\mu_1=0.830\rm{GeV},\mu_2=0.176\rm{GeV}$. In the left panel, from top to bottom, the lines stand for $B=0,0.025,0.05,0.075,0.100,0.125,0.150,0.175,0.200,0.23\rm{GeV}^2$ respectively. In the right panel, we show how $T_c$ changes with $B$. In low $B$, the errors due to the uncertainty of $\mu$ are quite small. In this range decreasing behavior of $T_c$ with increasing $B$ are reliable. In large $B$ region, full numerical solutions are necessary to get the correct behavior of $T_c$.}
\label{Tc-B-m0}
\end{figure}

Firstly, we check the chiral limit case by taking $m_q=0$, and extract $\sigma$ for different temperatures and magnetic fields. The numerical results are shown in Fig.\ref{sigma-B-T-m0}. In this figure, we plot the temperature dependent chiral condensate for $B=0,0.05,0.1,0.15,0.2,0.23{\rm GeV}^2$. In Panel.(a), we give the results for small $B$. For each value of $B$, we estimates the errors by varying $\mu$ from $0.25{\rm GeV}$ to $1 {\rm GeV}$. Then we see that at temperature range from $60{\rm MeV}$ to $160 {\rm MeV}$, the three lines for different value of $\mu$ almost overlap, which shows that in the case of small magnetic field, the perturbative expansion is good approximation at this temperature region. Also, we see that at the same temperature, the chiral condensate decreases when $B$ increases. Furthermore, we could see that for each $B$, a characteristic second order phase transition appears and the transition temperature decreases with increasing $B$, showing a kind of IMC effect. Then, in Panel.(b), we give the results for larger $B$. We could see that when $B$ is larger than $0.15{\rm GeV}^2$, the differences between lines for different values of $\mu$ become large even near the transition point, which confirms our analytical analysis that at large $B$ the perturbative expansion would not be good approximation anymore.
%

As can be seen in Fig.\ref{sigma-B-T-m0}, even with finite magnetic field, chiral phase transition remains second order in chiral limit. The phase transition point is well defined and we can extract the exact transition temperature easily by numerical calculation. So we also give the $\mu$ dependent transition temperature for several values of magnetic field in Fig.\ref{Tc-B-m0}. In Fig.\ref{Tc-B-m0}(a), we plot the $\mu$ dependent transition temperature for $B=0,0.025,0.05,0.075,0.100,0.125,0.150,$ $0.175,0.200,0.23\rm{GeV}^2$ from top to down respectively. From the plot, we could see that for small $B$(up to $B=0.15 {\rm GeV}^2$), the $\mu$ dependence is very weak. In some sense, it shows that the validity of the perturbative solution in this region. In addition, we could see an obvious decreasing behavior of transition temperature with increasing $B$. To be more explicitly, we plot the $\mu$ and $B$ dependent behavior of  $T_c$ in Panel.(b). The band in Panel.(b) comes from varying $\mu$ and could be considered as an estimate of the errors by using the perturbative expansion. In small $B$ region, we see that the band is very narrow, while in large $B$ it becomes wide. From Panel.(b), we can also see an obvious IMC effect in the small $B$ region. The qualitative behavior in the large $B$ region is unknown and needs further analysis. We will leave it in the future.

In a short summary, we find that in the range $B<0.15{\rm GeV}^2$(related to a physical magnetic field $\mathcal{B}\approx0.26{\rm GeV}^2$) the perturbative expansion could be good approximation even near the transition temperature. In this region, the numerical results using the perturbative expansion are reliable and it shows a characteristic IMC effect. The numerical results confirm our analysis from the behavior of $-\log(\sqrt{q}h)$ in last section, which might be related to the running behavior of coupling constant with magnetic field.

\subsection{Inverse magnetic catalysis at finite quark mass}

\begin{figure}[h]
\begin{center}
\epsfxsize=6.5 cm \epsfysize=6.5 cm \epsfbox{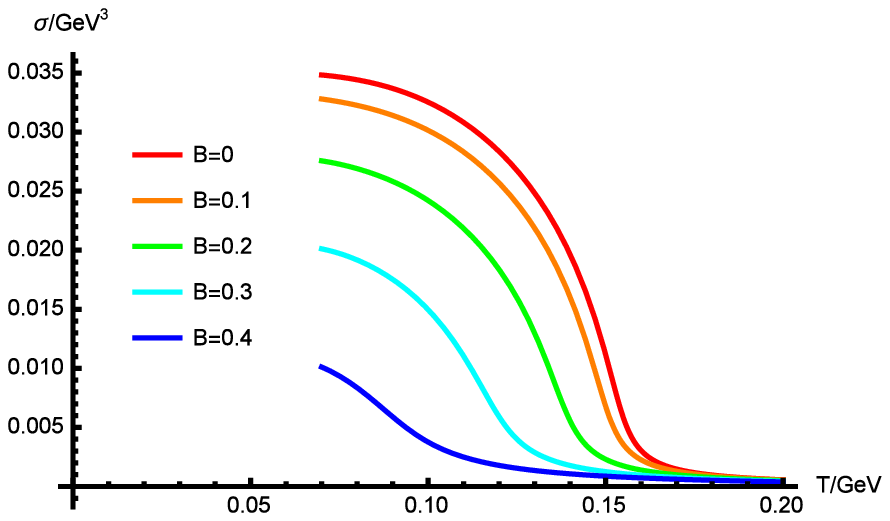}
\hspace*{0.1cm} \epsfxsize=6.5 cm \epsfysize=6.5 cm \epsfbox{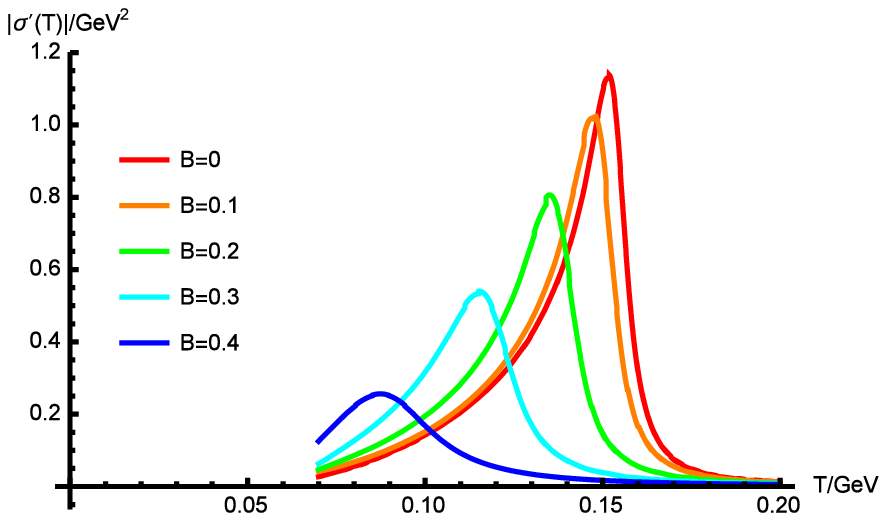}
\vskip -0.05cm \hskip 0.15 cm
\textbf{( a ) } \hskip 6.5 cm \textbf{( b )} \\
\end{center}
\caption{$\sigma$ and $\frac{d\sigma}{dT}$ as a function of temperature for different values of magnetic field $B=0,0.1,0.2,0.3,0.4{\rm GeV}^2$, when $m_q=5{\rm MeV},v_4=8,\mu_0=0.430\rm{GeV},\mu_1=0.830\rm{GeV},\mu_2=0.176\rm{GeV}$. Panel.(a) gives results of $\sigma$ for $B=0,0.1,0.2,0.3,0.4{\rm GeV}^2$ in red, orange, green, cyan, blue solid lines. Panel.(b) gives results of $\frac{d\sigma}{dT}$ for $B=0,0.1,0.2,0.3,0.4{\rm GeV}^2$ in red, orange, green, cyan, blue solid lines.}
\label{sigma-B-T-m5}
\end{figure}

\begin{figure}[h]
\begin{center}
\epsfxsize=6.5 cm \epsfysize=6.5 cm \epsfbox{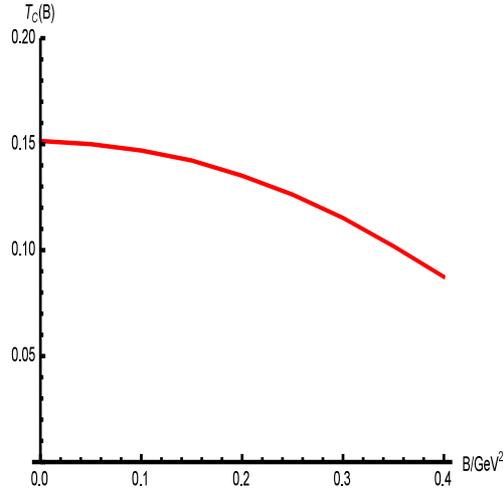}
\end{center}
\caption{Critical temperature $T_c$ as magnetic field $B$  when $m_q=5{\rm MeV},v_4=8,\mu_0=0.430\rm{GeV},\mu_1=0.830\rm{GeV},\mu_2=0.176\rm{GeV}$ and $\mu=1\rm{GeV}$. } \label{Tc-B-m5}
\end{figure}

It is also interesting to consider the case with finite quark mass, especially near physical point. Without loss of generality, we take $m_q=5{\rm MeV}$ as an example to show the qualitative behavior at finite quark mass. Since in last section we have seen that in the valid region of the approximated expansion, the $\mu$ dependence of the results are weak, we will fix $\mu=1{\rm GeV}$ in this section and only consider the results in small $B$ region as reliable results. Taking $m_q=5{\rm MeV}$ and solving the chiral condensate from Eq.(\ref{equation-chi}), we plot the results for different values of $B$ in Fig.\ref{sigma-B-T-m5}. In Fig.\ref{sigma-B-T-m5}(a), we give the temperature dependent chiral condensate for $B=0,0.1,0.2,0.3,0.4{\rm GeV}^2$ in red, orange, green, cyan, blue solid lines respectively.  From the lines, we could see that the chiral phase transition turns to be a crossover one at finite quark mass. In this case, there are no exact transition points. One possible way is to define the pseudo transition point at the location of the fastest transition rate of $\sigma$, where $|d\sigma/dT|$ takes its maximum value. To show the behavior of the pseudo transition point under finite magnetic field, in Fig.\ref{sigma-B-T-m5}(b) we also show the results of $|d\sigma/dT|$ for different $B$. From the figures, we can also read that the chiral condensate is suppressed by magnetic field at temperature near transition point, though from the perturbative solution we can not get the information for low temperature case. Then from Fig.\ref{sigma-B-T-m5}(b), we see that the location of the peak in $|d\sigma/dT|$ moves towards low temperature region when increasing the magnetic field. Extracting $T_c$ from Fig.\ref{sigma-B-T-m5}(b) and plotting them in Fig.\ref{Tc-B-m5}, we could see this effects more explicitly. There the pesudo transition temperature decreases with the increasing magnetic field. Though in Fig.\ref{Tc-B-m5} we plot the figure in a wide range of $B$, we should keep in mind that the reliable results are only in small $B$ range. In large $B$ range, it is still unknown whether the transition temperature will decrease or increase with $B$. To get the information, a full solution with magnetic field is required, which will be done in the near future.

Therefore, we have seen inverse magnetic catalysis effect in the modified soft-wall AdS/QCD model both in chiral limit and finite quark mass cases. Mathematically, it is mainly because the extra terms in the effective dilaton field $\phi$ is positive. We have checked the arguments in previous section numerically, and we emphasize that the positivity of this extra term is guaranteed by the Einstein-Maxwell action other than model dependent effect. The dilaton field could be dual to the coupling constant \cite{Gursoy:2008za,He:2010ye}. In some sense, the origin of IMC effect in the holographic scenario might be attribute to the correct running behavior of coupling constant, which is similar to the studies in Refs.\cite{Ferreira:2014kpa,Ferrer:2014qka}.

\section{Conclusion and discussion}
\label{sum}

The framework of soft-wall AdS/QCD has been widely tested from hadron physics and relevant phenomena. It is also very convenient to consider current quark mass, chiral condensate and chiral phase transition. Thus, it is very interesting to investigate magnetic effects on chiral phase transition within the framework of soft-wall AdS/QCD. The recent study in \cite{Dudal:2015wfn} shows that there are no inverse magnetic catalysis effect in the original soft-wall model. However, as pointed out in \cite{Chelabi:2015cwn,Chelabi:2015gpc}, the original soft-wall model with a pure quadratic dilaton filed can not describe chiral phase transition well. So we consider the modified model with a dilaton field negative at ultraviolet region and positive at infrared region and an extra quartic term in the scalar potential, which has been shown to give well description on chiral phase transition in \cite{Chelabi:2015cwn,Chelabi:2015gpc}.

To introduce magnetic field in the soft-wall model, we follow the strategy in \cite{Mamo:2015dea,Dudal:2015wfn} and consider the magnetized metric solved from Einstein-Maxwell system. We try to work out the magnetized background geometry from the Einstein-Maxwell system. Due to the complexity of the equation of motion, we use the leading perturbative expansion instead of the full solution. We show that when $B$ is very small, the perturbative expansion has good convergency and can be good approximation to the full solution, while when $B$ becomes large the perturbative expansion turns to be unreliable.

Then we embed the magnetized background metric into the modified soft-wall model. We solve the magnetic field dependent chiral condensate both in chiral limit and at finite quark mass. To estimate the errors by using the perturbative solution, we vary $\mu$ from $0.2 {\rm GeV}$ to $1 {\rm GeV}$. From Fig.\ref{sigma-B-T-m0}, we see that when $B$ is smaller than $0.15 {\rm GeV}^2$, the results would not depend on $\mu$ sensitively in the relevant region, showing that the perturbative solution could be considered as good approximation, while when $B$ is large it can not. In reliable region, we find that the phase transition remains a second order one in chiral limit, while at finite quark mass it turns to be a crossover one. Furthermore, near the transition temperature, we find that both in chiral limit and at finite quark mass chiral condensate is suppressed by magnetic field and the transition temperature would decrease with increasing of magnetic field, showing a kind of inverse magnetic catalysis effect. From an rough analytic analysis in Sec.\ref{magneticeff}, we could see the main reason for the IMC effect in the holographic model is that magnetic field causes an effective term $-\log(\sqrt{q}h)$ in the dilaton field $\phi$. Since the dilaton field could be dual to the coupling constant, this result might support the study in Refs.\cite{Ferreira:2014kpa,Ferrer:2014qka} that the effective running of the coupling constant with $B$ might cause the IMC effect.

In this work, we have not considered the back-reaction of dilaton field to the background metric, and the dilaton field does not depend on magnetic field. In principle, if we consider a more general Einstein-Dilaton-Maxwell system, we might get more consistent results. Also, since we use the perturbative expansion only, the large magnetic field $B$ and small temperature $T$ results are unreliable. We will leave the more consistent scenario and the full solution in the future.

\vskip 0.5cm
{\bf Acknowledgement}
\vskip 0.2cm
The authors thank Kaddour Chelabi, Zheng Fang, Song He and Yue-Liang Wu for valuable discussions. M.H. is supported by the NSFC under Grant Nos. 11175251 and 11621131001, DFG and NSFC (CRC 110), CAS key project KJCX2-EW-N01. D.L. is partly supported by China Postdoctoral Science Foundation(2015M580136). Y.Y. and P-H.Y. are supported by the Ministry of Science and Technology (NSC 101-2112-M-009-005).

\end{document}